\documentclass[preprint,superscriptaddress,nofootinbib]{revtex4}
  \usepackage{graphicx}
  \begin{document}
  \title{Study of ${\Upsilon}(nS)$ ${\to}$ $B_{c}P$ decays
         with perturbative QCD approach}
  \author{Yueling Yang}
  \affiliation{Institute of Particle and Nuclear Physics,
              Henan Normal University, Xinxiang 453007, China}
  \author{Junfeng Sun}
  \affiliation{Institute of Particle and Nuclear Physics,
              Henan Normal University, Xinxiang 453007, China}
  \author{Yan Guo}
  \affiliation{Institute of Particle and Nuclear Physics,
              Henan Normal University, Xinxiang 453007, China}
  \author{Qingxia Li}
  \affiliation{Institute of Particle and Nuclear Physics,
              Henan Normal University, Xinxiang 453007, China}
  \author{Jinshu Huang}
  \affiliation{College of Physics and Electronic Engineering,
              Nanyang Normal University, Nanyang 473061, China}
  \author{Qin Chang}
  \affiliation{Institute of Particle and Nuclear Physics,
              Henan Normal University, Xinxiang 453007, China}

  \begin{abstract}
  With the potential prospects of the ${\Upsilon}(nS)$
  at high-luminosity dedicated heavy-flavor factories,
  the color-favored ${\Upsilon}(nS)$ ${\to}$ $B_{c}{\pi}$,
  $B_{c}K$ weak decays are studied with the pQCD approach.
  It is found that branching ratios for the ${\Upsilon}(nS)$
  ${\to}$ $B_{c}{\pi}$ decay are as large as the order of
  ${\cal O}(10^{-11})$, which might be measured promisingly
  by the future experiments.
  \end{abstract}
  \pacs{13.25.Gv 12.39.St 14.40.Pq}
  \maketitle

  \section{Introduction}
  \label{sec01}
  Since the discovery of upsilons in proton-nucleus collisions
  at Fermilab in 1977 \cite{herb,innes},
  remarkable achievements have been made in the understanding
  of the nature of bottomonium (bound state of $b\bar{b}$).
  The upsilon ${\Upsilon}(nS)$ is the spin-triplet $S$-wave
  state $n^{3}S_{1}$ of bottomonium with the well established
  quantum number of $I^{G}J^{PC}$ $=$ $0^{-}1^{--}$ \cite{pdg}.
  The spectroscopy, production and decay mechanisms of the
  bottomonium resemble those of charmonium.
  The upsilon ${\Upsilon}(nS)$ below the $B\bar{B}$ threshold with
  the radial quantum number $n$ $=$ 1, 2 and 3 (note that for
  simplicity, the notation ${\Upsilon}(nS)$ will denote all
  ${\Upsilon}(1S)$, ${\Upsilon}(2S)$ and ${\Upsilon}(3S)$ mesons
  in the following content if not specified explicitly),
  in a close analogy with $J/{\psi}$,
  decay primarily through the annihilation of the $b\bar{b}$
  pairs into three gluons, followed by the evolution of gluons
  into hadrons, glueballs, multiquark and other exotic states.
  The strong ${\Upsilon}(nS)$ decay offers an ideal plaza to
  glean the properties of the invisible gluons and of the
  quark-gluon coupling \cite{ann1983}.
  The ${\Upsilon}(nS)$ strong decays are suppressed by the
  Okubo-Zweig-Iizuka rules \cite{o,z,i},
  which enable electromagnetic and radiative transitions
  to become competitive\footnotemark[1].
  \footnotetext[1]{Because of $G$-parity conservation,
  there also exist dipion transitions ${\Upsilon}(3S)$ ${\to}$
  ${\pi}{\pi}{\Upsilon}(2S)$, ${\pi}{\pi}{\Upsilon}(1S)$
  and ${\Upsilon}(2S)$ ${\to}$ ${\pi}{\pi}{\Upsilon}(1S)$,
  and hadronic transitions ${\Upsilon}(3S,2S)$ ${\to}$
  ${\eta}{\Upsilon}(1S)$ \cite{pdg,1212.6552}.}
  Besides, the upsilon weak decay is also legitimate
  within the standard model, although the branching ratio is
  tiny, about $2/{\tau}_{B}{\Gamma}_{\Upsilon}$ ${\sim}$
  ${\cal O}(10^{-8})$ \cite{pdg}.
  In this paper, we will estimate the branching ratios
  for the bottom-changing nonleptonic ${\Upsilon}(nS)$
  ${\to}$ $B_{c}P$ weak decays with perturbative QCD
  (pQCD) approach \cite{pqcd1,pqcd2,pqcd3}, where $P$
  denotes pseudoscalar ${\pi}$ and $K$ mesons.
  The motivation is listed as follows.

  From the experimental point of view,
  (1)
  over $10^{8}$ ${\Upsilon}(nS)$ samples have been accumulated
  at Belle and Babar collaborations due to their outstanding
  performance \cite{1406.6311} (see Table.\ref{tab:bb}).
  It is hopefully expected that more than $10^{11}$ $b\bar{b}$
  quark pairs would be available per $fb^{-1}$ data at LHCb \cite{1408.0403}.
  Much more upsilons could be collected with
  great precision at the forthcoming SuperKEKB and the running
  upgraded LHC, which provide a golden opportunity to search
  for the ${\Upsilon}(nS)$ weak decays that in some cases might
  be detectable.
  Theoretical studies of the ${\Upsilon}(nS)$ weak decays
  are very necessary to offer a ready reference.
  (2)
  For the two-body ${\Upsilon}(nS)$ ${\to}$ $B_{c}{\pi}$, $B_{c}K$ decays,
  the back-to-back final states with opposite charges have
  definite energies and momenta in the center-of-mass frame
  of upsilons.
  Additionally, identification of a single flavored $B_{c}$
  meson is free from inefficiently double tagging of flavored
  hadron pairs produced via conventional decays occurring
  above the $B\bar{B}$ threshold \cite{zpc62.271},
  and can also provide a conclusive evidence of the upsilon
  weak decay.
  Of course, small branching ratios make the observation
  of the upsilon weak decays extremely challenging,
  and the observation of an abnormally large production rate
  of single $B_{c}$ mesons in the ${\Upsilon}(nS)$ decay
  might be a hint of new physics \cite{zpc62.271}.

   \begin{table}[h]
   \caption{Summary of the mass, decay width and data samples
   of upsilon ${\Upsilon}(1S,2S,3S)$.}
   \label{tab:bb}
   \begin{ruledtabular}
   \begin{tabular}{ccccc}
    & \multicolumn{2}{c}{properties \cite{pdg}}
    & \multicolumn{2}{c}{data samples ($10^{6}$) \cite{1406.6311}} \\ \cline{2-3} \cline{4-5}
      meson & mass (MeV) & width (keV) & Belle & BaBar \\ \hline
   ${\Upsilon}(1S)$ & $9460.30{\pm}0.26$ & $54.02{\pm}1.25$
                    & $102{\pm}2$ & ...   \\
   ${\Upsilon}(2S)$ & $10023.26{\pm}0.31$ & $31.98{\pm}2.63$
                    & $158{\pm}4$ & $98.3{\pm}0.9$ \\
   ${\Upsilon}(3S)$ & $10355.2{\pm}0.5 $ & $20.32{\pm}1.85$
                    & $11{\pm}0.3$ &  $121.3{\pm}1.2$
   \end{tabular}
   \end{ruledtabular}
   \end{table}

  From the theoretical point of view,
  the bottom-changing upsilon weak decays permit one
  to reexamine parameters obtained from $B$ meson decay,
  test various phenomenological models and improve our
  understanding on the strong interactions and the
  mechanism responsible for heavy meson weak decay.
  The ${\Upsilon}(nS)$ ${\to}$ $B_{c}P$ decays are
  monopolized by tree contributions and favored by the
  Cabibbo-Kobayashi-Maskawa (CKM) matrix element
  $V_{cb}$, so they should have relatively large
  branching ratios among nonleptonic upsilon weak decays.
  The ${\Upsilon}(1S)$ ${\to}$ $B_{c}{\pi}$, $B_{c}K$ decays
  have been studied with the naive factorization
  (NF) approximation in previous works
  \cite{zpc62.271,ijma14,adv2013}.
  One obvious deficiency of NF approach is the
  disappearance of strong phases and the renormalization
  scale from hadronic matrix elements (HME).
  Recently, several attractive methods have been
  developed to reevaluate HME,
  such as pQCD \cite{pqcd1,pqcd2,pqcd3},
  the QCD factorization (QCDF) \cite{qcdf1,qcdf2,qcdf3}
  and soft and collinear effective theory
  \cite{scet1,scet2,scet3,scet4}.
  These methods have been widely used and could explain
  reasonably many measurements on nonleptonic $B_{u,d}$ decays.
  But, few works devote to the nonleptonic upsilon weak decays
  with these new phenomenological approaches.
  In this paper, we will study the ${\Upsilon}(nS)$ ${\to}$
  $B_{c}{\pi}$, $B_{c}K$ weak decays with the pQCD approach.

  This paper is organized as follows.
  In section \ref{sec02}, we present the theoretical framework
  and the amplitudes for the ${\Upsilon}(nS)$ ${\to}$ $B_{c}{\pi}$,
  $B_{c}K$ decays with pQCD approach.
  Section \ref{sec03} is devoted to numerical results and discussion.
  The last section is our summary.

  \section{theoretical framework}
  \label{sec02}
  \subsection{The effective Hamiltonian}
  \label{sec0201}
  The effective Hamiltonian for the
  ${\Upsilon}(nS)$ ${\to}$ $B_{c}{\pi}$, $B_{c}K$
  decays is written as \cite{9512380}
   \begin{equation}
  {\cal H}_{\rm eff}\ =\ \frac{G_{F}}{\sqrt{2}}\,
   \sum\limits_{q=d,s}\, V_{cb} V_{uq}^{\ast}\,
   \Big\{ C_{1}({\mu})\,Q_{1}({\mu})
         +C_{2}({\mu})\,Q_{2}({\mu}) \Big\}
   + {\rm h.c.}
   \label{hamilton},
   \end{equation}
  where $G_{F}$ is the Fermi coupling constant;
  the CKM factors are expanded as a power series in the small
  Wolfenstein parameter ${\lambda}$ ${\sim}$ 0.2 \cite{pdg},
  \begin{eqnarray}
  V_{cb}V_{ud}^{\ast} &=&
               A{\lambda}^{2}
  - \frac{1}{2}A{\lambda}^{4}
  - \frac{1}{8}A{\lambda}^{6}
  +{\cal O}({\lambda}^{8})
  \label{eq:ckm01}, \\
  V_{cb}V_{us}^{\ast} &=& A{\lambda}^{3}
  +{\cal O}({\lambda}^{8})
  \label{eq:ckm02}.
  \end{eqnarray}
  The Wilson coefficients $C_{1,2}(\mu)$ summarize the
  physical contributions above scales of ${\mu}$,
  and have been properly calculated to the NLO order
  with the renormalization group improved perturbation
  theory.
  The local operators are defined as follows.
    \begin{eqnarray}
    Q_{1} &=&
  [ \bar{c}_{\alpha}{\gamma}_{\mu}(1-{\gamma}_{5})b_{\alpha} ]
  [ \bar{q}_{\beta} {\gamma}^{\mu}(1-{\gamma}_{5})u_{\beta} ]
    \label{q1}, \\
    Q_{2} &=&
  [ \bar{c}_{\alpha}{\gamma}_{\mu}(1-{\gamma}_{5})b_{\beta} ]
  [ \bar{q}_{\beta}{\gamma}^{\mu}(1-{\gamma}_{5})u_{\alpha} ]
    \label{q2},
    \end{eqnarray}
  where ${\alpha}$ and ${\beta}$ are color indices and the
  sum over repeated indices is understood.

  \subsection{Hadronic matrix elements}
  \label{sec0202}
  To obtain the decay amplitudes, one has to calculate the
  hadronic matrix elements of local operators.
  Analogous to the common applications of hard exclusive
  processes in perturbative QCD proposed by Lepage and
  Brodsky \cite{prd22}, HME
  could be expressed as the convolution of hard scattering
  subamplitudes containing perturbative contributions
  with the universal wave functions reflecting the
  nonperturbative contributions.
  However, sometimes, the high-order corrections to HME
  produce collinear and/or soft logarithms
  based on collinear factorization approximation,
  for example, the spectator scattering amplitudes
  within the QCDF framework \cite{qcdf3}.
  The pQCD approach advocates
  that \cite{pqcd1,pqcd2,pqcd3} this problem could be
  settled down by retaining the transverse momentum of
  quarks and introducing the Sudakov factor.
  The decay amplitudes could be factorized into three
  parts: the soft effects uniting with the universal
  wave functions ${\Phi}$, the process-dependent
  subamplitude $H$, the hard effects incorporated
  into the Wilson coefficients $C_{i}$.
  For a particular topology,
  the decay amplitudes could be written as
  \begin{equation}
  {\cal A}_{i}\ {\propto}\ \prod\limits_{j}
  {\int}dx_{j}\,db_{j}\, C_{i}(t)\,H_{i}(t_{i},x_{j},b_{j})\,
  {\Phi}_{j}(x_{j},b_{j})\,e^{-S}
  \label{hadronic},
  \end{equation}
  where $t_{i}$ is a typical scale, $x_{j}$ is the
  longitudinal momentum fraction of the valence quark,
  $b_{j}$ is the conjugate variable of the transverse
  momentum, $e^{-S}$ is the Sudakov factor, and
  $j$ denotes participating particles.

  \subsection{Kinematic variables}
  \label{sec0203}
  In the ${\Upsilon}(nS)$ rest frame, the light cone kinematic
  variables are defined as
  \begin{equation}
  p_{{\Upsilon}}\, =\, p_{1}\, =\, \frac{m_{1}}{\sqrt{2}}(1,1,0)
  \label{kine-p1},
  \end{equation}
  \begin{equation}
  p_{B_{c}}\, =\, p_{2}\, =\, (p_{2}^{+},p_{2}^{-},0)
  \label{kine-p2},
  \end{equation}
  \begin{equation}
  p_{\pi(K)}\, =\, p_{3}\, =\, (p_{3}^{-},p_{3}^{+},0)
  \label{kine-p3},
  \end{equation}
  \begin{equation}
  k_{i}\, =\, x_{i}\,p_{i}+(0,0,\vec{k}_{i{\perp}})
  \label{kine-ki},
  \end{equation}
  \begin{equation}
  {\epsilon}_{\Upsilon}^{\parallel}\, =\, \frac{1}{ \sqrt{2} }(1,-1,0)
  \label{kine-1el},
  \end{equation}
  \begin{equation}
  n_{+}=(1,0,0), \quad n_{-}=(0,1,0)
  \label{kine-null},
  \end{equation}
  \begin{equation}
  p_{i}^{\pm}\, =\, \frac{E_{i}\,{\pm}\,p}{\sqrt{2}}
  \label{kine-pipm},
  \end{equation}
  \begin{equation}
  p\, =\, \frac{ \sqrt{ [m_{1}^{2}-(m_{2}+m_{3})^{2}]
                        [m_{1}^{2}-(m_{2}-m_{3})^{2}] }}{2\,m_{1}}
  \label{kine-pcm},
  \end{equation}
  \begin{equation}
  s\, =\, 2\,p_{2}{\cdot}p_{3}
   \, =\, m_{1}^{2}-m_{2}^{2}-m_{3}^{2}
  \label{kine-s},
  \end{equation}
  \begin{equation}
  t\, =\, 2\,p_{1}{\cdot}p_{2}
   \, =\, m_{1}^{2}+m_{2}^{2}-m_{3}^{2}
   \, =\, 2\,m_{1}\,E_{2}
  \label{kine-t},
  \end{equation}
  \begin{equation}
  u\, =\, 2\,p_{1}{\cdot}p_{3}
   \, =\, m_{1}^{2}-m_{2}^{2}+m_{3}^{2}
   \, =\, 2\,m_{1}\,E_{3}
  \label{kine-u},
  \end{equation}
  \begin{equation}
  t+u-s\, =\, m_{1}^{2}+m_{2}^{2}+m_{3}^{2}
  \label{kine-stu01},
  \end{equation}
  where $x_{i}$ and $\vec{k}_{i{\perp}}$ are the longitudinal momentum
  fraction and transverse momentum of the light valence quark, respectively;
  ${\epsilon}_{\Upsilon}^{\parallel}$ is the longitudinal polarization vector
  of the ${\Upsilon}(nS)$ particle;
  $n_{+}$ and $n_{-}$ are positive and negative null vectors, respectively;
  $p$ is the common momentum of final states;
  $m_{1}$, $m_{2}$ and $m_{3}$ denote the masses of the ${\Upsilon}(nS)$,
  $B_{c}$ and ${\pi}(K)$ mesons, respectively.
  The notation of momentum is displayed in Fig.\ref{fig:amp}(a).

  \subsection{Wave functions}
  \label{sec0204}
  With the notation in
  \cite{prd65.014007,npb529.323,jhep0605.004},
  the definitions of matrix elements of
  diquark operators sandwiched between
  vacuum and the longitudinally polarized
  ${\Upsilon}(nS)$,
  the double-heavy pseudoscalar $B_{c}$,
  the light pseudoscalar $P$ are
  \begin{equation}
 {\langle}0{\vert}b_{i}(z)\bar{b}_{j}(0){\vert}
 {\Upsilon}(p_{1},{\epsilon}_{\parallel}){\rangle}\,
 =\, \frac{1}{4}f_{\Upsilon}
 {\int}dx_{1}\,e^{-ix_{1}p_{1}{\cdot}z}
  \Big\{ \!\!\not{\epsilon}_{\parallel} \Big[
   m_{1}\,{\phi}_{\Upsilon}^{v}(x_{1})
  -\!\!\not{p}_{1}\, {\phi}_{\Upsilon}^{t}(x_{1})
  \Big] \Big\}_{ji}
  \label{wave-bbl},
  \end{equation}
  \begin{equation}
 {\langle}B_{c}^{+}(p_{2}){\vert}\bar{c}_{i}(z)b_{j}(0){\vert}0{\rangle}\,
 =\, \frac{i}{4}f_{B_{c}} {\int}dx_{2}\,e^{ix_{2}p_{2}{\cdot}z}\,
  \Big\{ {\gamma}_{5}\Big[ \!\!\not{p}_{2}\, {\phi}_{B_{c}}^{a}(x_{2})
  +m_{2}\,{\phi}_{B_{c}}^{p}(x_{2}) \Big] \Big\}_{ji}
  \label{wave-bc2},
  \end{equation}
  \begin{eqnarray}
  & &
 {\langle}P(p_{3}){\vert}u_{i}(z)\bar{q}_{j}(0){\vert}0{\rangle}
  \nonumber \\ &=&
  \frac{i}{4}f_{P} {\int}dx_{3}\,e^{ix_{3}p_{3}{\cdot}z}
  \Big\{ {\gamma}_{5}\Big[ \!\!\not{p}_{3}\,{\phi}_{P}^{a}(x_{3})
  + {\mu}_{P}{\phi}_{P}^{p}(x_{3})
  - {\mu}_{P}(\!\not{n}_{-}\!\!\not{n}_{+}\!-\!1)\,{\phi}_{P}^{t}(x_{3})
  \Big] \Big\}_{ji}
  \label{wave-pim},
  \end{eqnarray}
  where $f_{\Upsilon}$, $f_{B_{c}}$, $f_{P}$ are
  decay constants,
  ${\mu}_{P}$ $=$ $m_{3}^{2}/(m_{u}+m_{q})$ and
  $q$ $=$ $d(s)$ for ${\pi}(K)$ meson.

  The twist-2 distribution amplitudes of light pseudoscalar
  ${\pi}$, $K$ mesons are defined as \cite{jhep0605.004}:
   \begin{equation}
  {\phi}_{P}^{a}(x)=6\,x\bar{x}
   \Big\{ 1+ \sum\limits_{n=1}^{\infty}
   a_{n}^{P}\, C_{n}^{3/2}(x-\bar{x}) \Big\}
   \label{twist},
   \end{equation}
  where $\bar{x}$ $=$ $1$ $-$ $x$;
  $a_{n}^{P}$ and $C_{n}^{3/2}(z)$ are
  Gegenbauer moment and polynomials, respectively;
  $a^{\pi}_{i}$ $=$ $0$ for $i$ $=$ 1, 3, 5, ${\cdots}$
  due to the explicit $G$-parity of pion.

  Both ${\Upsilon}(nS)$ and $B_{c}$ systems are nearly
  nonrelativistic, due to $m_{{\Upsilon}}$ ${\simeq}$
  $2m_{b}$ and $m_{B_{c}}$ ${\simeq}$ $m_{b}$ $+$ $m_{c}$.
  Nonrelativistic quantum chromodynamics (NRQCD)
  \cite{prd46,prd51,rmp77} and
  Schr\"{o}dinger equation can be used to describe
  their spectrum.
  The radial wave functions with isotropic harmonic
  oscillator potential are written as
   \begin{equation}
  {\phi}_{1S}(\vec{k})\
  {\sim}\ e^{-\vec{k}^{2}/2{\beta}^{2}}
   \label{wave-p-1s},
   \end{equation}
   \begin{equation}
  {\phi}_{2S}(\vec{k})\ {\sim}\
   e^{-\vec{k}^{2}/2{\beta}^{2}}
   ( 2\vec{k}^{2}-3{\beta}^{2} )
   \label{wave-p-r2s},
   \end{equation}
   \begin{equation}
  {\phi}_{3S}(\vec{k})\ {\sim}\
   e^{-\vec{k}^{2}/2{\beta}^{2}}
   ( 4\vec{k}^{4}-20\vec{k}^{2}{\beta}^{2}+15{\beta}^{4} )
   \label{wave-p-r3s},
   \end{equation}
  where the parameter ${\beta}$ determines the average
  transverse momentum, i.e.,
  ${\langle}1S{\vert}\vec{k}^{2}_{\perp}{\vert}1S{\rangle}$
  $=$ ${\beta}^{2}$.
  According to the NRQCD power counting rules \cite{prd46},
  the characteristic magnitude of the momentum of heavy quark
  is order of $Mv$, where $M$ is the mass of the heavy quark
  with typical velocity $v$ ${\sim}$ ${\alpha}_{s}(M)$.
  So, value of ${\beta}$ $=$ $M{\alpha}_{s}(M)$ is taken
  in our calculation.
  Employing the substitution ansatz \cite{xiao},
   \begin{equation}
   \vec{k}^{2}\ {\to}\ \frac{1}{4} \sum\limits_{i}
   \frac{\vec{k}_{i\perp}^{2}+m_{q_{i}}^{2}}{x_{i}}
   \label{wave-kt},
   \end{equation}
  where $x_{i}$, $\vec{k}_{i\perp}$, $m_{q_{i}}$ are the
  longitudinal momentum fraction, transverse momentum,
  mass of the light valence quark, respectively,
  with the relations ${\sum}x_{i}$ $=$ $1$ and
  $\sum\vec{k}_{i\perp}$ $=$ $0$.
  Integrating out $\vec{k}_{i\perp}$ and combining with
  their asymptotic forms, one can obtain
   \begin{equation}
  {\phi}_{B_{c}}^{a}(x) = A\, x\bar{x}\,
  {\exp}\Big\{ -\frac{\bar{x}\,m_{c}^{2}+x\,m_{b}^{2}}
                     {8\,{\beta}_{2}^{2}\,x\,\bar{x}} \Big\}
   \label{wave-bca},
   \end{equation}
   \begin{equation}
  {\phi}_{B_{c}}^{p}(x) = B\,
  {\exp}\Big\{ -\frac{\bar{x}\,m_{c}^{2}+x\,m_{b}^{2}}
                     {8\,{\beta}_{2}^{2}\,x\,\bar{x}} \Big\}
   \label{wave-bcp},
   \end{equation}
   \begin{equation}
  {\phi}_{{\Upsilon}(1S)}^{v}(x) = C\, x\bar{x}\,
  {\exp}\Big\{ -\frac{m_{b}^{2}}{8\,{\beta}_{1}^{2}\,x\,\bar{x}} \Big\}
   \label{wave-bbv},
   \end{equation}
   \begin{equation}
  {\phi}_{{\Upsilon}(1S)}^{t}(x) = D\, (x-\bar{x})^{2}\,
  {\exp}\Big\{ -\frac{m_{b}^{2}}{8\,{\beta}_{1}^{2}\,x\,\bar{x}} \Big\}
   \label{wave-bbt},
   \end{equation}
   \begin{equation}
  {\phi}_{{\Upsilon}(2S)}^{t,v}(x) = E\,
  {\phi}_{{\Upsilon}(1S)}^{t,v}(x)\,
   \Big\{ 1+\frac{m_{b}^{2}}{2\,{\beta}_{1}^{2}\,x\,\bar{x}} \Big\}
   \label{wave-bb2s},
   \end{equation}
   \begin{equation}
  {\phi}_{{\Upsilon}(3S)}^{t,v}(x) = F\,
  {\phi}_{{\Upsilon}(1S)}^{t,v}(x)\,
   \Big\{ \Big( 1-\frac{m_{b}^{2}}{2\,{\beta}_{1}^{2}\,x\,\bar{x}} \Big)^{2}
   +6 \Big\}
   \label{wave-bb3s},
   \end{equation}
   where ${\beta}_{i}$ $=$ ${\xi}_{i}{\alpha}_{s}({\xi}_{i})$
   with ${\xi}_{i}$ $=$ $m_{i}/2$;
   parameters $A$, $B$, $C$, $D$, $E$, $F$ are the normalization
   coefficients satisfying the conditions
   \begin{equation}
  {\int}_{0}^{1}dx\,{\phi}_{B_{c}}^{a,p}(x)=1,
   \quad
  {\int}_{0}^{1}dx\,{\phi}_{\Upsilon}^{v,t}(x) =1
   \label{wave-abc}.
   \end{equation}

\begin{figure}[h]
  \includegraphics[width=0.9\textwidth,bb=75 570 540 720]{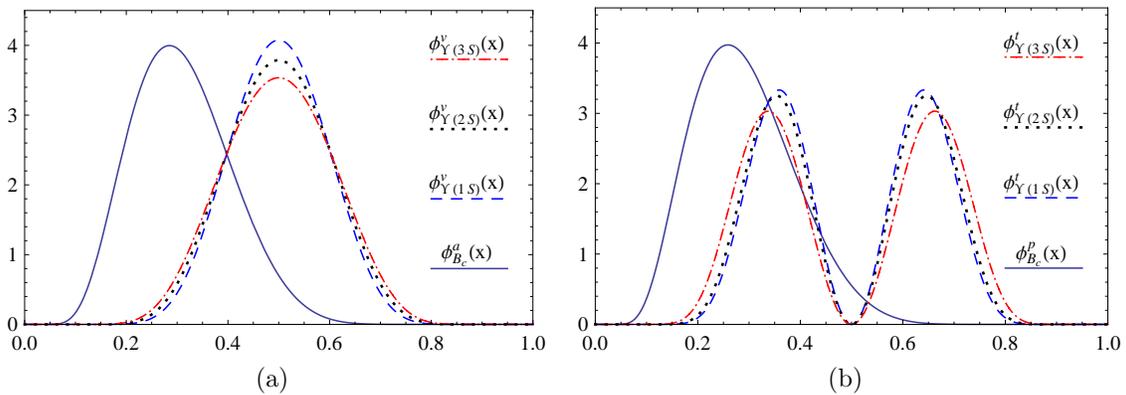}
  \caption{The distribution amplitudes of
  ${\phi}_{B_{c}}^{a,p}(x)$ and ${\phi}_{{\Upsilon}(nS)}^{v,t}(x)$.}
  \label{fig:wf}
  \end{figure}

  The shape lines of the normalized distribution amplitudes of
  ${\phi}_{B_{c}}^{a,p}(x)$ and ${\phi}_{{\Upsilon}(nS)}^{v,t}(x)$
  are displayed in Fig.\ref{fig:wf}.
  Here we would like to point out that the relativistic
  corrections of ${\cal O}(v^{2})$ are left out.
  According to the arguments in Ref. \cite{prd46},
  the ${\cal O}(v^{2})$ corrections could bring about
  10${\sim}$30\% errors, and it is expected that such
  error could be reduced systematically by including new
  interactions in principle, which is beyond the scope
  of this paper.

  \subsection{Decay amplitudes}
  \label{sec0205}
\begin{figure}[h]
  \includegraphics[width=0.99\textwidth,bb=75 620 530 720]{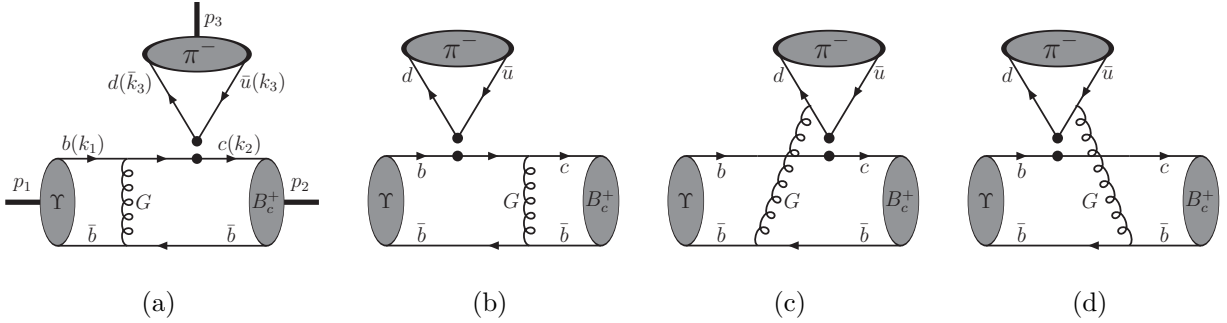}
  \caption{Feynman diagrams for the ${\Upsilon}(nS)$ ${\to}$ $B_{c}{\pi}$
   decay with the pQCD approach.}
  \label{fig:amp}
  \end{figure}

  The Feynman diagrams for the ${\Upsilon}(nS)$ ${\to}$ $B_{c}{\pi}$
  decay within the pQCD framework are shown in Fig.\ref{fig:amp},
  where (a) and (b) are factorizable topology;
  (c) and (d) are nonfactorizable topology.

  The decay amplitudes of ${\Upsilon}(nS)$ ${\to}$ $B_{c}P$
  decay can be written as
   \begin{equation}
  {\cal A}({\Upsilon}(nS){\to}B_{c}P)
   = \sqrt{2}G_{F}\frac{{\pi}C_{F}}{N} V_{cb} V_{uq}^{\ast}\,
   m_{\Upsilon}^{3}\,pf_{\Upsilon}f_{B_{c}}f_{P}\!\!\!\!
   \sum\limits_{i=a,b,c,d}\!\!\!\!
  {\cal A}_{\rm Fig.\ref{fig:amp}(i)}
   \label{amp-all},
   \end{equation}
  where $C_{F}$ $=$ $4/3$ and the color number $N$ $=$ $3$.

  The explicit expressions of
  ${\cal A}_{\rm Fig.\ref{fig:amp}(i)}$
  are
   \begin{eqnarray}
   \lefteqn{ {\cal A}_{\rm Fig.\ref{fig:amp}(a)}\, =\,
  {\int}_{0}^{1}dx_{1} {\int}_{0}^{\infty}b_{1} db_{1}
  {\int}_{0}^{1}dx_{2} {\int}_{0}^{\infty}b_{2} db_{2}\,
  {\alpha}_{s}(t_{a})\, a_{1}(t_{a})\, E_{a}(t_{a}) }
   \nonumber \\ & & {\times}
  H_{a}(x_{1},x_{2},b_{1},b_{2})\,
  {\phi}_{\Upsilon}^{v}(x_{1})\, \Big\{
  {\phi}_{B_{c}}^{a}(x_{2})\,(x_{2}+r_{3}^{2}\,\bar{x}_{2})
 +{\phi}_{B_{c}}^{p}(x_{2})\,r_{2}\,r_{b} \Big\}
   \label{amp-figa},
   \end{eqnarray}
   \begin{eqnarray}
   \lefteqn{ {\cal A}_{\rm Fig.\ref{fig:amp}(b)}\, =\,
  {\int}_{0}^{1}dx_{1} {\int}_{0}^{\infty}b_{1} db_{1}
  {\int}_{0}^{1}dx_{2} {\int}_{0}^{\infty}b_{2} db_{2}\,
  {\alpha}_{s}(t_{b})\, a_{1}(t_{b})\, E_{b}(t_{b}) }
   \nonumber \\ & & {\times}
   H_{b}(x_{1},x_{2},b_{2},b_{1})\,
   \Big\{ 2\,r_{2}\, {\phi}_{B_{c}}^{p}(x_{2}) \Big[
    r_{c}\, {\phi}_{\Upsilon}^{v}(x_{1})
   +x_{1}\, {\phi}_{\Upsilon}^{t}(x_{1}) \Big]
   \nonumber \\ & &
  -{\phi}_{B_{c}}^{a}(x_{2}) \Big[
   {\phi}_{\Upsilon}^{v}(x_{1})\,
   (r_{2}^{2}\,x_{1}+ r_{3}^{2}\,\bar{x}_{1})
  +{\phi}_{\Upsilon}^{t}(x_{1})\, r_{c} \Big] \Big\}
   \label{amp-figb},
   \end{eqnarray}
   \begin{eqnarray}
   \lefteqn{ {\cal A}_{\rm Fig.\ref{fig:amp}(c)}\, =\,
  {\int}_{0}^{1}dx_{1} {\int}_{0}^{\infty} db_{1}
  {\int}_{0}^{1}dx_{2} {\int}_{0}^{\infty}b_{2} db_{2}
  {\int}_{0}^{1}dx_{3} {\int}_{0}^{\infty}b_{3} db_{3}\,
  {\delta}(b_{1}-b_{2})\,{\alpha}_{s}(t_{c}) }
   \nonumber \\ & & {\times}
   \frac{C_{2}(t_{c})}{N}\,
   E_{c}(t_{c})\, H_{c}(x_{1},x_{2},x_{3},b_{2},b_{3})\,
   {\phi}_{P}^{a}(x_{3})\, \Big\{ {\phi}_{B_{c}}^{p}(x_{2})\,
  {\phi}_{\Upsilon}^{t}(x_{1})\, r_{2}\,(x_{2}-x_{1})
   \nonumber \\ & & \qquad +\
  {\phi}_{B_{c}}^{a}(x_{2})\,
  {\phi}_{\Upsilon}^{v}(x_{1})\,
   \Big[ (1+r_{2}^{2}-r_{3}^{2})\,(x_{1}-x_{3})
   +2\,r_{2}^{2}\,(x_{3}-x_{2}) \Big] \Big\}
   \label{amp-figc},
   \end{eqnarray}
   \begin{eqnarray}
   \lefteqn{ {\cal A}_{\rm Fig.\ref{fig:amp}(d)}\, =\,
  {\int}_{0}^{1}dx_{1} {\int}_{0}^{\infty} db_{1}
  {\int}_{0}^{1}dx_{2} {\int}_{0}^{\infty}b_{2} db_{2}
  {\int}_{0}^{1}dx_{3} {\int}_{0}^{\infty}b_{3} db_{3}\,
  {\delta}(b_{1}-b_{2})\, {\alpha}_{s}(t_{d})}
   \nonumber \\ & & {\times}
  \frac{C_{2}(t_{d})}{N}\,
  E_{d}(t_{d})\, H_{d}(x_{1},x_{2},x_{3},b_{2},b_{3})\,
  {\phi}_{P}^{a}(x_{3})\, \Big\{ {\phi}_{B_{c}}^{p}(x_{2})\,
  {\phi}_{\Upsilon}^{t}(x_{1})\, r_{2}\,(x_{2}-x_{1})
   \nonumber \\ & & \qquad +\
  {\phi}_{B_{c}}^{a}(x_{2})\, {\phi}_{\Upsilon}^{v}(x_{1})\,
  (1-r_{2}^{2}-r_{3}^{2})\,(\bar{x}_{2}-x_{3}) \Big\}
   \label{amp-figd},
   \end{eqnarray}
  where ${\alpha}_{s}$ is the QCD coupling;
  $a_{1}$ $=$ $C_{1}$ $+$ $C_{2}/N$;
  $C_{1,2}$ is the Wilson coefficients;
  $r_{i}$ $=$ $m_{i}/m_{1}$.
  It can be easily seen that (1) the nonfactorizable
  contributions ${\cal A}_{\rm Fig.\ref{fig:amp}(c,d)}$
  are color-suppressed with respect to the
  factorizable contributions
  ${\cal A}_{\rm Fig.\ref{fig:amp}(a,b)}$;
  (2) The twist-3 distribution amplitudes ${\phi}_{P}^{p,t}$
  have no contribution to decay amplitudes.

  The typical scales $t_{i}$ and the Sudakov factor $E_{i}$
  are defined as
   \begin{equation}
   t_{a(b)} = {\max}(\sqrt{-{\alpha}_{g}},
                  \sqrt{-{\beta}_{a(b)}},
                  1/b_{1},1/b_{2})
   \label{tab},
   \end{equation}
   \begin{equation}
   t_{c(d)} = {\max}(\sqrt{-{\alpha}_{g}},
                  \sqrt{{\vert}{\beta}_{c(d)}{\vert}},
                  1/b_{1},1/b_{2},1/b_{3})
   \label{tcd},
   \end{equation}
   \begin{equation}
   E_{a(b)}(t) = {\exp}\{-S_{B_{c}}(t) \}
   \label{eab},
   \end{equation}
   \begin{equation}
   E_{c(d)}(t) = {\exp}\{-S_{B_{c}}(t)-S_{P}(t) \}
   \label{ecd},
   \end{equation}
   \begin{equation}
  {\alpha}_{g} = \bar{x}_{1}^{2}m_{1}^{2}
               +  \bar{x}_{2}^{2}m_{2}^{2}
               -  \bar{x}_{1}\bar{x}_{2}t
   \label{tg},
   \end{equation}
   \begin{equation}
  {\beta}_{a} = m_{1}^{2} - m_{b}^{2}
              +  \bar{x}_{2}^{2}m_{2}^{2}
              -  \bar{x}_{2}t
   \label{tqa},
   \end{equation}
   \begin{equation}
  {\beta}_{b} = m_{2}^{2} - m_{c}^{2}
              +  \bar{x}_{1}^{2}m_{1}^{2}
              -  \bar{x}_{1}t
   \label{tqb},
   \end{equation}
   \begin{equation}
  {\beta}_{c} = x_{1}^{2}m_{1}^{2}
              +  x_{2}^{2}m_{2}^{2}
              +  x_{3}^{2}m_{3}^{2}
              -  x_{1}x_{2}t
              -  x_{1}x_{3}u
              +  x_{2}x_{3}s
   \label{tqc},
   \end{equation}
   \begin{equation}
  {\beta}_{d} = \bar{x}_{1}^{2}m_{1}^{2}
              +  \bar{x}_{2}^{2}m_{2}^{2}
              +  x_{3}^{2}m_{3}^{2}
              -  \bar{x}_{1}\bar{x}_{2}t
              -  \bar{x}_{1}x_{3}u
              +  \bar{x}_{2}x_{3}s
   \label{tqd},
   \end{equation}
   \begin{equation}
   S_{B_{c}}(t) = s(x_{2},p_{2}^{+},1/b_{2})
                +2{\int}_{1/b_{2}}^{t}\frac{d{\mu}}{\mu} {\gamma}_{q}
   \label{sbc},
   \end{equation}
   \begin{equation}
   S_{P}(t) = s(x_{3},p_{3}^{+},1/b_{3})
            + s(\bar{x}_{3},p_{3}^{+},1/b_{3})
            +2{\int}_{1/b_{3}}^{t}\frac{d{\mu}}{\mu}{\gamma}_{q}
   \label{sp},
   \end{equation}
  where ${\alpha}_{g}$ and ${\beta}_{i}$ are the virtuality
  of the internal gluon and quark, respectively;
  ${\gamma}_{q}$ $=$ $-{\alpha}_{s}/{\pi}$ is the quark
  anomalous dimension; the expression of $s(x,Q,1/b)$ can
  be found in the appendix of Ref.\cite{pqcd1}.

  The scattering functions $H_{i}$ in the subamplitudes
  ${\cal A}_{\rm Fig.\ref{fig:amp}(i)}$
  are defined as
   \begin{equation}
   H_{a(b)}(x_{1},x_{2},b_{i},b_{j})\, =\,
   K_{0}(\sqrt{-{\alpha}_{g}}b_{i})
   \Big\{ {\theta}(b_{i}-b_{j})
   K_{0}(\sqrt{-{\beta}}b_{i})
   I_{0}(\sqrt{-{\beta}}b_{j})
   + (b_{i}{\leftrightarrow}b_{j}) \Big\}
   \label{hab},
   \end{equation}
   \begin{eqnarray}
   H_{c(d)}(x_{1},x_{2},x_{3},b_{2},b_{3}) &=&
   \Big\{ {\theta}(-{\beta}) K_{0}(\sqrt{-{\beta}}b_{3})
  +\frac{{\pi}}{2} {\theta}({\beta}) \Big[
   iJ_{0}(\sqrt{{\beta}}b_{3})
   -Y_{0}(\sqrt{{\beta}}b_{3}) \Big] \Big\}
   \nonumber \\ &{\times}&
   \Big\{ {\theta}(b_{2}-b_{3})
   K_{0}(\sqrt{-{\alpha}_{g}}b_{2})
   I_{0}(\sqrt{-{\alpha}_{g}}b_{3})
   + (b_{2}{\leftrightarrow}b_{3}) \Big\}
   \label{hcd},
   \end{eqnarray}
  where $J_{0}$ and $Y_{0}$ ($I_{0}$ and $K_{0}$) are the
  (modified) Bessel function of the first and second kind,
  respectively.

  \section{Numerical results and discussion}
  \label{sec03}

  In the rest frame of the ${\Upsilon}(nS)$ particle,
  branching ratio for the ${\Upsilon}(nS)$ ${\to}$
  $B_{c}P$ weak decays can be written as
   \begin{equation}
  {\cal B}r({\Upsilon}(nS){\to}B_{c}P)\ =\ \frac{1}{12{\pi}}\,
   \frac{p}{m_{{\Upsilon}}^{2}{\Gamma}_{{\Upsilon}}}\,
  {\vert}{\cal A}({\Upsilon}(nS){\to}B_{c}P){\vert}^{2}
   \label{br}.
   \end{equation}

   \begin{table}[h]
   \caption{Numerical values of the input parameters}
   \label{tab:input}
   \begin{ruledtabular}
   \begin{tabular}{ll}
   Wolfenstein parameters \cite{pdg}:
   & $A$ $=$ $0.814^{+0.023}_{-0.024}$,\quad
     ${\lambda}$ $=$ $0.22537{\pm}0.00061$; \\ \hline
   Masses of quarks \cite{pdg} :
   & $m_{c}$ $=$ $1.67{\pm}0.07$ GeV, \quad
     $m_{b}$ $=$ $4.78{\pm}0.06$ GeV; \\ \hline
   Gegenbauer moments:
   & $a_{2}^{\pi}$ (1 GeV) $=$ $0.17{\pm}0.08$ \cite{prd83},\quad
     $a_{4}^{\pi}$ (1 GeV) $=$ $0.06{\pm}0.10$ \cite{prd83},\\
   & $a_{1}^{K}$ (1 GeV) $=$ $0.06{\pm}0.03$ \cite{jhep0605.004},\quad
     $a_{2}^{K}$ (1 GeV) $=$ $0.25{\pm}0.15$ \cite{jhep0605.004}; \\ \hline
   decay constant:
   & $f_{\pi}$ $=$ $130.41{\pm}0.20$ MeV \cite{pdg}, \quad
     $f_{K}$ $=$ $156.2{\pm}0.7$ MeV \cite{pdg},\\
   & $f_{B_{c}}$ $=$ $489{\pm}5$ MeV \cite{fbc}, \quad
     $f_{{\Upsilon}(1S)}$ $=$ $(676.4{\pm}10.7)$ MeV, \\
   & $f_{{\Upsilon}(2S)}$ $=$ $(473.0{\pm}23.7)$ MeV, \quad
     $f_{{\Upsilon}(3S)}$ $=$ $(409.5{\pm}29.4)$ MeV.
   \end{tabular}
   \end{ruledtabular}
   \end{table}

  The input parameters are collected in Table. \ref{tab:input}.
  As for the decay constant $f_{{\Upsilon}}$, one can use
  the definition of decay constant,
   \begin{equation}
  {\langle}0{\vert}\bar{b}{\gamma}^{\mu}{b}{\vert}{\Upsilon}{\rangle}
   = f_{{\Upsilon}}m_{{\Upsilon}}{\epsilon}_{{\Upsilon}}^{\mu}
   \label{fvv1}.
   \end{equation}
  and relate $f_{{\Upsilon}}$ to the experimentally
  measurable leptonic branching ratio,
   \begin{equation}
  {\Gamma}({\Upsilon}{\to}{\ell}^{+}{\ell}^{-})\, =\,
   \frac{4{\pi}}{27}{\alpha}_{\rm QED}^{2}
   \frac{f_{{\Upsilon}}^{2}}{m_{{\Upsilon}}}
   \sqrt{ 1-2\frac{m_{\ell}^{2}}{m_{{\Upsilon}}^{2}} }
   \Big\{ 1+2\frac{m_{\ell}^{2}}{m_{{\Upsilon}}^{2}} \Big\}
   \label{fvv2},
   \end{equation}
  where ${\alpha}_{\rm QED}$ is the fine-structure
  constant, $m_{\ell}$ is the lepton mass and ${\ell}$
  $=$ $e$, ${\mu}$, ${\tau}$.

  The values of $f_{{\Upsilon}}$ determined from measurements
  are listed in Table.\ref{tab:fbb}.
  One may notice that there are some clear hierarchical
  relations among these decay constant.
  (1) The decay constants
  $f_{{\Upsilon}(nS)}^{e^{+}e^{-}}$ $<$
  $f_{{\Upsilon}(nS)}^{{\mu}^{+}{\mu}^{-}}$ $<$
  $f_{{\Upsilon}(nS)}^{{\tau}^{+}{\tau}^{-}}$ for
  the same radial quantum number $n$.
  There are two reasons.
  One is that the final phase space for decay into
  ${\ell}_{i}^{+}{\ell}_{i}^{-}$ states is more compact
  than that for ${\ell}_{j}^{+}{\ell}_{j}^{-}$ decay
  when the lepton family number $i$ $>$ $j$.
  The other is that branching ratio of ${\ell}_{i}^{+}{\ell}_{i}^{-}$
  decay is relatively less than that of ${\ell}_{j}^{+}{\ell}_{j}^{-}$
  decay with the lepton family number $i$ $<$ $j$.
  (2) There are also two reasons for the relation among the
  weighted average $f_{{\Upsilon}(1S)}$ $>$ $f_{{\Upsilon}(2S)}$
  $>$ $f_{{\Upsilon}(3S)}$.
  One is that the possible phase space increases with the radial
  quantum number of upsilon due to $m_{{\Upsilon}(1S)}$ $<$
  $m_{{\Upsilon}(2S)}$ $<$ $m_{{\Upsilon}(3S)}$.
  The other is that decay width of upsilon decreases with
  the radial quantum number $n$.

   \begin{table}[h]
   \caption{Branching ratios for leptonic ${\Upsilon}(nS)$
   decays and decay constants $f_{{\Upsilon}}$, where
   the last column is the weighted average,
   and errors come from mass, width and
   branching ratios.}
   \label{tab:fbb}
   \begin{ruledtabular}
   \begin{tabular}{lccc}
   decay mode & branching ratio &
   \multicolumn{2}{c}{decay constant} \\ \hline
    ${\Upsilon}(1S)$ ${\to}$ $e^{+}e^{-}$
  & $(2.38{\pm}0.11)\%$
  & $(664.2{\pm}23.1)$ MeV & \\
    ${\Upsilon}(1S)$ ${\to}$ ${\mu}^{+}{\mu}^{-}$
  & $(2.48{\pm}0.05)\%$
  & $(677.9{\pm}14.7)$ MeV
  & $(676.4{\pm}10.7)$ MeV \\
    ${\Upsilon}(1S)$ ${\to}$ ${\tau}^{+}{\tau}^{-}$
  & $(2.60{\pm}0.10)\%$
  & $(683.3{\pm} 21.1)$ MeV & \\ \hline
    ${\Upsilon}(2S)$ ${\to}$ $e^{+}e^{-}$
  & $(1.91{\pm}0.16)\%$
  & $(471.0{\pm}39.1)$ MeV & \\
    ${\Upsilon}(2S)$ ${\to}$ ${\mu}^{+}{\mu}^{-}$
  & $(1.93{\pm}0.17)\%$
  & $(473.5{\pm}40.3)$ MeV
  & $(473.0{\pm}23.7)$ MeV \\
    ${\Upsilon}(2S)$ ${\to}$ ${\tau}^{+}{\tau}^{-}$
  & $(2.00{\pm}0.21)\%$
  & $(475.2{\pm} 44.5)$ MeV & \\ \hline
    ${\Upsilon}(3S)$ ${\to}$ ${\mu}^{+}{\mu}^{-}$
  & $(2.18{\pm}0.21)\%$
  & $(407.6{\pm}38.2)$ MeV
  & $(409.5{\pm}29.4)$ MeV \\
    ${\Upsilon}(3S)$ ${\to}$ ${\tau}^{+}{\tau}^{-}$
  & $(2.29 {\pm}0.30 )\%$
  & $(412.2 {\pm}45.9)$ MeV &
  \end{tabular}
  \end{ruledtabular}
  \end{table}
   \begin{table}[h]
   \caption{Branching ratios for the ${\Upsilon}(nS)$ ${\to}$
   $B_{c}{\pi}$, $B_{c}K$ decays, where previous results
   are calculated with the coefficient $a_{1}$ $=$ $1.05$.}
   \label{tabbr}
   \begin{ruledtabular}
  \begin{tabular}{lcccc}
    & Ref. \cite{ijma14} & Ref. \cite{adv2013}
    & Ref. \cite{691261} & this work \\ \hline
    $10^{11}{\times}{\cal B}r({\Upsilon}(1S){\to}B_{c}{\pi})$
  & 6.91 & 2.8 & 5.03
  & $7.40^{+0.51+0.90+0.88+0.30}_{-0.49-0.57-1.01-0.28}$ \\
    $10^{11}{\times}{\cal B}r({\Upsilon}(2S){\to}B_{c}{\pi})$
  & ... & ... & ...
  & $6.29^{+0.43+0.70+0.55+1.67}_{-0.41-0.45-0.62-0.44}$ \\
    $10^{11}{\times}{\cal B}r({\Upsilon}(3S){\to}B_{c}{\pi})$
  & ... & ... & ...
  & $6.57^{+0.45+0.69+0.44+1.55}_{-0.43-0.45-0.51-0.86}$ \\ \hline
    $10^{12}{\times}{\cal B}r({\Upsilon}(1S){\to}B_{c}K)$
  & 5.03 & 2.3 & 3.73
  & $5.67^{+0.42+0.71+0.68+0.24}_{-0.40-0.45-0.77-0.23}$ \\
    $10^{12}{\times}{\cal B}r({\Upsilon}(2S){\to}B_{c}K)$
  & ... & ... & ...
  & $4.85^{+0.36+0.55+0.45+1.27}_{-0.34-0.35-0.42-0.36}$ \\
    $10^{12}{\times}{\cal B}r({\Upsilon}(3S){\to}B_{c}K)$
  & ... & ... & ...
  & $5.09^{+0.38+0.55+0.35+1.15}_{-0.36-0.35-0.34-0.71}$
  \end{tabular}
  \end{ruledtabular}
  \end{table}

  Our numerical results on the $CP$-averaged branching ratios
  for the ${\Upsilon}(nS)$ ${\to}$ $B_{c}{\pi}$, $B_{c}K$ decays
  are displayed in Table \ref{tabbr},
  where the uncertainties come from the CKM parameters,
  the renormalization scale ${\mu}$ $=$ $(1{\pm}0.1)t_{i}$,
  masses of $b$ and $c$ quarks, hadronic parameters including
  decay constants and Gegenbauer moments, respectively.
  The following are some comments.

  (1)
  Branching ratios for the bottom-changing ${\Upsilon}(nS)$
  ${\to}$ $B_{c}{\pi}$, $B_{c}K$ weak decays with the pQCD
  approach have the same magnitude of order as previous
  estimation in Refs. \cite{ijma14,adv2013,691261}.
  Compared with the NF and QCDF approaches,
  there are more contributions from the nonfactorizable
  decay amplitudes ${\cal A}_{\rm Fig.\ref{fig:amp}(c,d)}$
  with the pQCD approach, which may be the reason of why
  the pQCD's results are slightly larger than previous ones.

  (2)
  Because of hierarchical relation between the CKM factors
  ${\vert}V_{cb}V_{us}^{\ast}{\vert}$ $>$
  ${\vert}V_{cb}V_{ud}^{\ast}{\vert}$, in general,
  there is relation between branching ratios
  ${\cal B}r({\Upsilon}(nS){\to}B_{c}{\pi})$ $>$
  ${\cal B}r({\Upsilon}(nS){\to}B_{c}K)$.

  (3)
  Because the relations among masses $m_{{\Upsilon}(3S)}$
  $>$ $m_{{\Upsilon}(2S)}$ $>$ $m_{{\Upsilon}(1S)}$ resulting
  in that the momentum and phase space of final states increase
  with the radial quantum number $n$, in addition, the relation
  among decay widths ${\Gamma}_{{\Upsilon}(3S)}$ $<$
  ${\Gamma}_{{\Upsilon}(2S)}$ $<$ ${\Gamma}_{{\Upsilon}(1S)}$
  (see Table. \ref{tab:bb}),
  in principle, it is expected that there should be relations
  among branching ratios ${\cal B}r({\Upsilon}(3S){\to}B_{c}P)$
  $>$ ${\cal B}r({\Upsilon}(2S){\to}B_{c}P)$ $>$
  ${\cal B}r({\Upsilon}(1S){\to}B_{c}P)$ for the same
  pseudoscalar meson $P$.
  But the results in Table. \ref{tabbr} is not the way
  one expected it to be. Why? The reason is that the decay
  amplitudes with the pQCD approach is proportional to
  decay constant $f_{{\Upsilon}(nS)}$, and the fact of
  that the difference among final phase spaces is small,
  hence there is an approximation,
   \begin{eqnarray}
   & &
   {\cal B}r({\Upsilon}(1S){\to}B_{c}P) :
   {\cal B}r({\Upsilon}(2S){\to}B_{c}P) :
   {\cal B}r({\Upsilon}(3S){\to}B_{c}P)
   \nonumber \\ &{\propto}&
   \frac{ f_{{\Upsilon}(1S)}^{2} }{ {\Gamma}_{{\Upsilon}(1S)} } :
   \frac{ f_{{\Upsilon}(2S)}^{2} }{ {\Gamma}_{{\Upsilon}(2S)} } :
   \frac{ f_{{\Upsilon}(3S)}^{2} }{ {\Gamma}_{{\Upsilon}(3S)} }\
   {\simeq}\
   1.2 : 1 : 1.2
   \label{f2g}.
   \end{eqnarray}

  (3)
  Branching ratio for the ${\Upsilon}(nS)$ ${\to}$ $B_{c}{\pi}$
  decay is a few times of $10^{-11}$. So the nonleptonic
  ${\Upsilon}(nS)$ ${\to}$ $B_{c}{\pi}$ weak decays could be
  sought for with some priority at the running LHC and the
  forthcoming SuperKEKB.
  For example, the production cross section of ${\Upsilon}(nS)$
  in p-Pb collision can reach up to a few ${\mu}b$
  with the LHCb \cite{jhep1407} and ALICE \cite{plb740}
  detectors at LHC.
  Over $10^{11}$ ${\Upsilon}(nS)$ particles per 100 $fb^{-1}$
  data collected at LHCb and ALICE are in principle available,
  which corresponds to a few tens of ${\Upsilon}(nS)$ ${\to}$
  $B_{c}{\pi}$ events.

  (4)
  There are many uncertainties on our results.
  The CKM factors can bring about 7\% uncertainty on
  the prediction of branching ratio.
  More than 10\% uncertainty come from the variation of
  typical scale $t_{i}$.
  The effects of masses of $m_{b}$ and $m_{c}$ on
  branching ratio decrease with the radial quantum
  number $n$.
  Compared with the ${\Upsilon}(1S)$ weak decays,
  hadronic parameters give a noticeable uncertainty
  on ${\Upsilon}(2S,3S)$ weak decays due to large
  errors on the decay constants $f_{{\Upsilon}(2S,3S)}$
  relative to $f_{{\Upsilon}(1S)}$.
  Other factors, such as the contributions of higher
  order corrections to HME, relativistic effects
  and so on, which are not considered here,
  deserve the dedicated study.
  Our results just provide an order of magnitude estimation.

  \section{Summary}
  \label{sec04}
  The ${\Upsilon}(nS)$ weak decay is allowable within
  the standard model, although branching ratio is tiny
  compared with the strong and electromagnetic decays.
  It is expected that the ${\Upsilon}(nS)$ particles could
  be produced and collected copiously at the high-luminosity
  dedicated heavy-flavor factories.
  It seems to have a good opportunity and a realistic
  possibility to search for the ${\Upsilon}(nS)$ weak
  decay experimentally.
  In this paper, we study the color-favored bottom-changing
  ${\Upsilon}(nS)$ ${\to}$ $B_{c}{\pi}$, $B_{c}K$ weak decays
  with the pQCD approach just to offer a ready reference
  to experimental analysis.
  It is found that branching ratios for the
  ${\Upsilon}(nS)$ ${\to}$ $B_{c}{\pi}$ decays are
  the order of ${\cal O}(10^{-11})$, which might be
  detectable in future experiments.

  \section*{Acknowledgments}
  We thank Professor Dongsheng Du (IHEP@CAS) and Professor
  Yadong Yang (CCNU) for helpful discussion.

  

\begin{thebibliography}{99}
  \bibitem{herb}
          S. Herb {\em et al.}, Phys. Rev. Lett. 39, 252 (1977).
  \bibitem{innes}
          W. Innes {\em et al.}, Phys. Rev. Lett. 39, 1240 (1977).
  \bibitem{pdg}
          K. Olive {\em et al.} (Particle Data Group), Chin. Phys. C 38, 090001 (2014).
  \bibitem{ann1983}
          P. Franzini, J. Lee-Franzini, Ann. Rev. Nucl. Part. Sci. 33, 1 (1983).
  \bibitem{o}
          S. Okubo, Phys. Lett. 5, 165 (1963).
  \bibitem{z}
          G. Zweig, CERN-TH-401, 402, 412 (1964).
  \bibitem{i}
          J. Iizuka, Prog. Theor. Phys. Suppl. 37-38, 21 (1966).
  \bibitem{1212.6552}
          C. Patrignani, T. Pedlar, J. Rosner, Annu. Rev. Nucl. Part. Sci. 63, 21 (2013).
  \bibitem{pqcd1}
          H. Li, Phys. Rev. D 52, 3958 (1995).
  \bibitem{pqcd2}
          C. Chang,  H. Li, Phys. Rev. D 55, 5577 (1997).
  \bibitem{pqcd3}
          T. Yeh, H. Li, Phys. Rev. D 56, 1615 (1997).
  \bibitem{1406.6311}
          Ed. A. Bevan {\em et al.}, Eur. Phys. J. C 74, 3026, (2014).
  \bibitem{1408.0403}
          T. Gershon, M. Needham, C. R. Physique 16, 435 (2015).
  \bibitem{zpc62.271}
          M. Sanchis-Lozano, Z. Phys. C 62, 271 (1994).
  \bibitem{ijma14}
          K. Sharma, R. Verma, Int. J. Mod. Phys. A 14, 937 (1999).
  \bibitem{adv2013}
          R. Dhir, R. Verma, A. Sharma, Adv. in High Energy Phys. 2013, 706543 (2013).
  \bibitem{qcdf1}
          M. Beneke {\em et al.}, Phys. Rev. Lett. 83, 1914 (1999).
  \bibitem{qcdf2}
          M. Beneke {\em et al.}, Nucl. Phys. B 591, 313 (2000).
  \bibitem{qcdf3}
          M. Beneke {\em et al.}, Nucl. Phys. B 606, 245 (2001).
  \bibitem{scet1}
          C. Bauer {\em et al.},  Phys. Rev. D 63, 114020 (2001).
  \bibitem{scet2}
          C. Bauer, D. Pirjol, I. Stewart, Phys. Rev. D 65, 054022 (2002).
  \bibitem{scet3}
          C. Bauer {\em et al.}, Phys. Rev. D 66, 014017 (2002).
  \bibitem{scet4}
          M. Beneke {\em et al.}, Nucl. Phys. B 643, 431 (2002).
  \bibitem{9512380}
          G. Buchalla, A. Buras, M. Lautenbacher, Rev. Mod. Phys. 68, 1125, (1996).
  \bibitem{prd22}
          G. Legage, S. Brodsky, Phys. Rev. D 22, 2157 (1980).
  \bibitem{prd65.014007}
          T. Kurimoto, H. Li, A. Sanda, Phys. Rev. D 65, 014007 (2001).
  \bibitem{npb529.323}
          P. Ball, V. Braun, Y. Koike, K. Tanaka, Nucl. Phys. B 529, 323 (1998).
  \bibitem{jhep0605.004}
          P. Ball, V. Braun, A. Lenz, JHEP, 0605, 004, (2006).
  \bibitem{prd46}
          G. Legage {\em et al.}, Phys. Rev. D 46, 4052 (1992).
  \bibitem{prd51}
          G. Bodwin, E. Braaten, G. Legage, Phys. Rev. D 51, 1125 (1995).
  \bibitem{rmp77}
          N. Brambilla {\em et al.}, Rev. Mod. Phys. 77, 1423 (2005).
  \bibitem{xiao}
          B. Xiao, X. Qin, B. Ma, Eur. Phys. J. A 15, 523 (2002).
  \bibitem{prd83}
          A. Khodjamirian, Th. Mannel, N. Offen, Y. Wang, Phys. Rev. D 83, 094031 (2011).
  \bibitem{fbc}
          T. Chiu, T. Hsieh, C. Huang, K. Ogawa, Phys. Lett. B 651, 171 (2007).
  \bibitem{691261}
          J. Sun {\em et al.}, Adv. in High Energy Phys. 2015, 691261 (2015).
  \bibitem{jhep1407}
          R. Aaij {\em et al.} (LHCb Collaboration), JHEP 1407, 094 (2014).
  \bibitem{plb740}
          B. Abelev {\em et al.} (ALICE Collaboration), Phys. Lett. B 740, 105 (2015).
  \end{thebibliography}
  \end{document}